\documentclass[aip,jap,reprint,preprintnumbers]{revtex4-1}
\usepackage{graphicx}
\usepackage{amsmath,amssymb,dcolumn,bm}

\begin{document}

\title{Laboratory hard X-ray photoelectron spectroscopy of
La$_{\bm{1-x}}$Sr$_{\bm{x}}$MnO$_{\bm{3}}$}

\author{Tomoko~Hishida}
\email{t-hishida@mg.ngkntk.co.jp}
\affiliation{NGK SPARK PLUG Co., Ltd., Komaki, Aichi 485-8510, Japan}

\author{Masaaki~Kobata}
\email{m-kobata@spring8.or.jp}
\affiliation{National Institute for Materials Science (NIMS), Sayo, Hyogo 679-5198, Japan}
\affiliation{Quantum Beam Science Directorate, Japan Atomic Energy Agency, Sayo, Hyogo 679-5148, Japan}

\author{Eiji~Ikenaga}
\affiliation{Japan Synchrotron Radiation Research Institute,
Sayo, Hyogo 679-5198, Japan}

\author{Takeharu~Sugiyama}
\affiliation{Research Center for Synchrotron Light Applications, Kyushu University, Kasugai,
Fukuoka 816-8505, Japan}

\author{Kazushige~Ohbayashi}
\affiliation{NGK SPARK PLUG Co., Ltd., Komaki, Aichi 485-8510, Japan}

\author{Keisuke~Kobayashi}
\affiliation{National Institute for Materials Science (NIMS), Sayo, Hyogo 679-5198, Japan}
\affiliation{Hiroshima Synchrotron Radiation Center, Hiroshima University, Higashihiroshima,
Hiroshima 739-0046, Japan}

\author{Mario~Okawa}
\affiliation{Department of Applied Physics, Tokyo University of Science,
Katsushika, Tokyo 125-8585, Japan}

\author{Tomohiko~Saitoh}
\email{t-saitoh@rs.kagu.tus.ac.jp}
\affiliation{Department of Applied Physics, Tokyo University of Science,
Katsushika, Tokyo 125-8585, Japan}

\preprint{Journal-ref.: Jpn. J. Appl. Phys. \textbf{54}, 083201 (2015)}

\begin{abstract}
A laboratory hard X-ray photoelectron spectroscopy (HXPS) system equipped with a monochromatic
Cr K$\alpha$ ($h\nu = 5414.7$ eV) X-ray source was applied to an investigation of the core-level
electronic structure of La$_{1-x}$Sr$_x$MnO$_3$.
No appreciable high binding-energy shoulder in the O $1s$ HXPS spectra were observed while an enhanced
low binding-energy shoulder structure in the Mn $2p_{3/2}$ HXPS spectra were observed, both of which
are manifestation of high bulk sensitivity.
Such high bulk sensitivity enabled us to track the Mn $2p_{3/2}$ shoulder structure in the full range of $x$,
giving us a new insight into the binding-energy shift of the Mn $2p_{3/2}$ core level.
Comparisons with the results using the conventional laboratory XPS ($h\nu = 1486.6$ eV) as well as
those using a synchrotron radiation source ($h\nu = 7939.9$ eV) demonstrate that HXPS is a powerful
and convenient tool to analyze the bulk electronic structure of a host of different compounds.
\end{abstract}

\maketitle

\section{Introduction}

X-ray photoelectron spectroscopy (XPS) is a widely-used technique for investigating the electronic
structure of solids.
Nevertheless, the relatively high surface sensitivity of conventional laboratory-based XPS constitutes
an essential disadvantage for this technique.
High surface sensitivity sometimes makes information on the bulk electronic states difficult to obtain,
because the results are influenced by large surface contributions such as inherent modifications of
the near surface regions and/or surface contaminations.
Hard X-ray photoelectron spectroscopy (HX-PES) which makes use of monochromatized undulator
hard X-rays, was first reported in 2003. It has allowed us to overcome this limitation.\cite{ref1,ref2}
This method uses photoelectrons with large kinetic energy, thereby considerably increasing the probin
depth, which minimizes or even eliminates the surface contributions in many cases.
This technique greatly expands the range of applications of photoelectron spectroscopy and offers
unique opportunities for investigating advanced functional materials such as nano-scale multilayers and
nano clusters.
For general users, however, HX-PES based on synchrotron radiation is often inconvenient, especially for
researchers who conduct most of their research at their own laboratories.
In addition, competition is so high for synchrotron beamtime that not enough proposals are accepted.
This situation can be a serious barrier preventing prompt analysis of exotic, newly synthesized materials.
Many of these problems are resolved by laboratory hard X-ray photoelectron spectroscopy (HXPS)
by using ``the high-energy angle-resolved photoelectron spectrometer for laboratory use (HEARP Lab.)''
developed recently by Kobata et al., which employs monochromatic Cr K$\alpha$ X-rays ($h\nu = 5414.7$
eV).
They demonstrated that the system provides practical throughput and energy resolution and applied it
to various materials to exhibit its versatility as a bulk-sensitive probe of the electronic and chemical
states of materials.\cite{ref3,ref4}

We used HXPS to investigate the electronic structure of a perovskite-type manganese oxide
La$_{1−x}$Sr$_x$MnO$_3$ (LSMO).
In spite of a long history since its first report,\cite{ref5,ref6} LSMO is still a well-known modern compound
because of colossal magnetoresistance (CMR).\cite{ref7,ref8,ref9,ref10,ref11}
This remarkable phenomenon is based on the unique electronic structure of a Mn$^{3+}$--Mn$^{4+}$
mixed-valence system with $O_h$ symmetry, which leads to the double-exchange (DE) mechanism.\cite{ref12} 
However, DE alone cannot explain the \textit{colossal} MR effects,\cite{ref13,ref14,ref15} so further study
of the electronic-structure of these materials is still required.
In this paper, we report a comprehensive comparative study of the CMR manganite (LSMO) based on
the conventional XPS (Al K$\alpha$, $h\nu = 1486.6$ eV), HXPS (Cr K$\alpha$, $h\nu = 5414.7$ eV),
and HX-PES ($h\nu = 7939.9$ eV).

\section{Experimental procedure}

Polycrystalline samples of La$_{1−x}$Sr$_x$MnO$_3$ ($x$ = 0.0, 0.1, 0.2, 0.33, 0.4, 0.5, 0.55, 0.67,
0.8, 0.9, 1.0) were prepared by solid-state reaction.
Starting powders of La(OH)$_3$ (99.9{\%}, Shin-Etsu Chemical), SrCO$_3$ (99.8{\%}, Sakai Chemical
Industry), and Mn$_2$O$_3$ (99.9{\%}, Kojundo Chemcal Laboratory) were weighed in specific
proportions and then mixed by ball milling in ethanol for 15 h with zirconia balls.
The mixtures were dried and heated at $1100^\circ$C for 2 h in air.
After cooling, the powders were crushed by the ball milling again for 15 h.
The powders were again dried, and then pressed into pellet form by applying an isostatic pressure of 0.8
ton/cm$^2$.
Finally, they were sintered at $1600^\circ$C for 1 h in air.

HXPS spectra were acquired by using ``HEARP Lab.'' system consisting of a VG-Scienta R4000 electron
analyzer and a monochromatic Cr K$\alpha$ X-ray source ($h\nu = 5414.7$ eV).
The energy resolution was about 0.55 eV. For all measurements, the X-ray beam was focused onto the
sample at a 100 $\mu$m $\phi$ spot.
The binding energy was calibrated based on the position of the Au $4f_{7/2}$ peak (84.0 eV) and
the Fermi level ($E_\text{F}$).
For the measurements, the samples were kept in ultrahigh vacuum below $1 \times 10^{-7}$ Torr.
XPS measurements were made with a PHI Quantera SXM instrument (base pressure $5 \times 10^{-9}$
Torr), which uses a monochromatic Al K$\alpha$ X-ray source ($h\nu = 1486.6$ eV) with an energy
resolution of about 0.64 eV full width at half maximum (FWHM).
The analyzer pass energy was set to 55 eV for narrow scans.
The binding energy was corrected by using the value of 84.0 eV from the Au $4f_{7/2}$ core-level spectrum.
The measurement vacuum was better than $1 \times 10^{-8}$ Torr.
HX-PES measurements using monochromatized synchrotron radiation ($h\nu = 7939.9$ eV) were also
carried out at the BL47XU undulator beamline of SPring-8.
The energy resolution was 0.29 eV FWHM.
The binding energy was calibrated by using the location of the Au $4f_{7/2}$ peak (84.0 eV) and
$E_\text{F}$.
To obtain fresh, clean surfaces, the samples were fractured in situ at room temperature just before each
measurement.
All measurements were made at room temperature.

\section{Results and Discussion}

\begin{figure}[t]
	\centering
	\includegraphics[width=86mm]{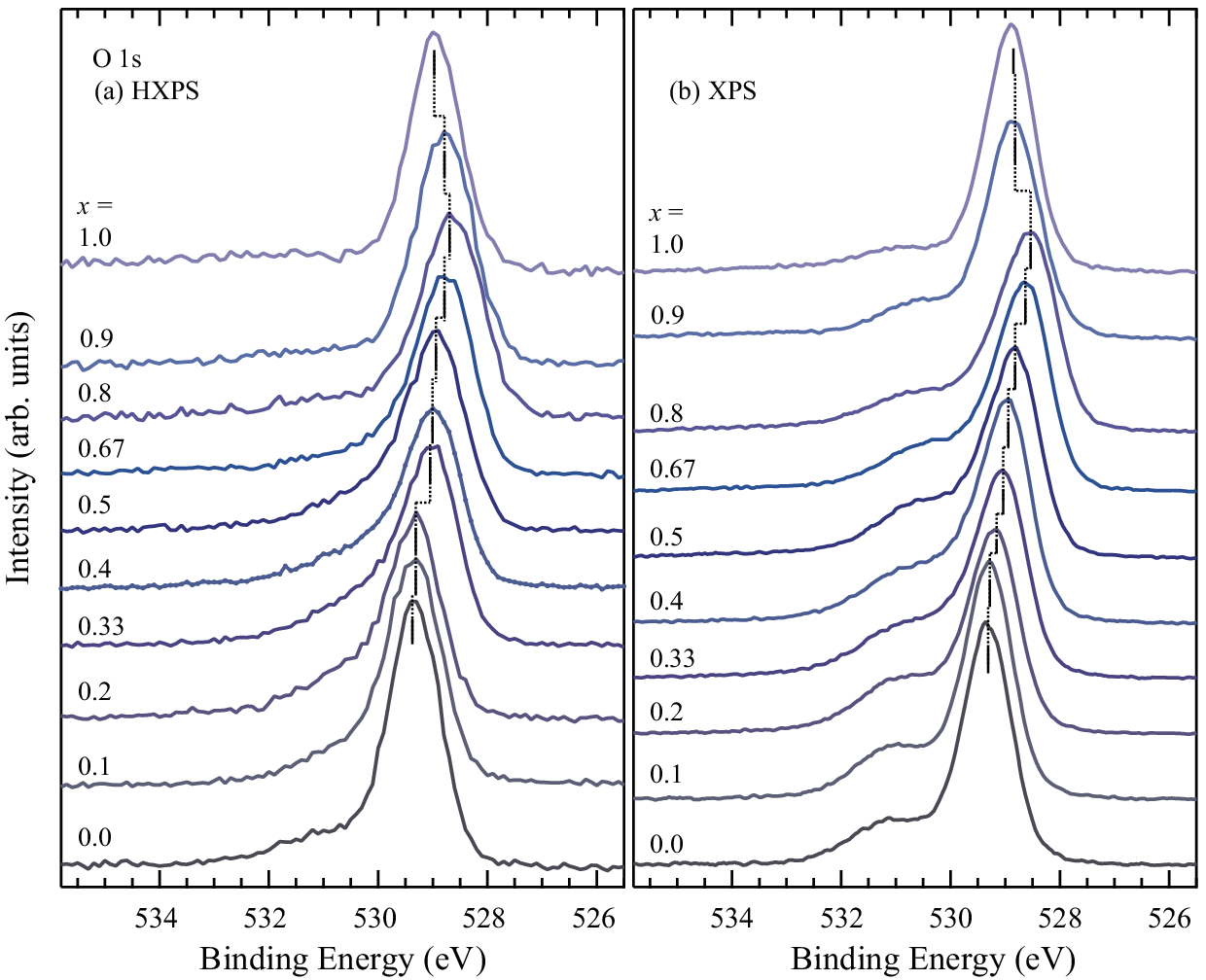}
	\caption{
		O $1s$ core-level photoelectron spectra of LSMO for various Sr concentrations,
		which were acquired by (a) HXPS and (b) XPS.
	}
	\label{fig1}
\end{figure}

Figures \ref{fig1}(a) and \ref{fig1}(b) compare the O $1s$ core-level spectra of LSMO measured
with HXPS and XPS, respectively.
The small hump near 531 eV (on the higher-binding-energy side of the main peak at $\sim$529 eV)
observed in the XPS spectra is attributed to a contribution from the sample surface.\cite{ref16,ref17,ref18}
The pectral weight of the hump is comparable to that reported by Bindu et al.\ and thus can be
estimated to be less than a few percent of that of the main peak.\cite{ref18}
The greatly improved bulk sensitivity of HXPS almost eliminates this hump and allows us to observe
a clear Doniach-\u{S}unji\'{c}-type asymmetric lineshape\cite{ref19} in the metallic range in $x$.

\begin{figure}[b]
	\centering
	\includegraphics[width=75mm]{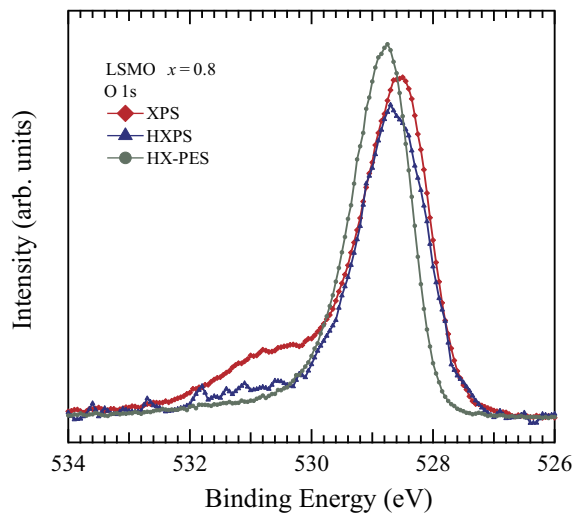}
	\caption{
		Comparison of O $1s$ core-level photoelectron spectra for $x=0.8$ acquired by
		XPS, HXPS, and HX-PES.
	}
	\label{fig2}
\end{figure}

Figure \ref{fig2} compares the O $1s$ core-level spectra of the $x = 0.8$ sample acquired by XPS,
HXPS, and HX-PES.
With respect to the XPS spectra, the HXPS (HX-PES) spectrum shifts towards higher energy by
0.09 eV (0.21 eV).
This shift is attributed to the recoil effect of oxygen atom in the photoemission process.
It is known that the energy shift $\Delta E_\text{R}$ due to the recoil effect of a free atom is
given by\cite{ref20,ref21}
\begin{equation*}
	\Delta E_\text{R} = \frac{q^2}{2M} = E_\text{k} \times \frac{m}{M}, \quad
	E_\text{k} = \frac{q^2}{2m} = h\nu,
\end{equation*}
where $q$ and $E_\text{k}$ are the momentum and the kinetic energy, respectively, of a
photo-emitted electron, and $M$ and $m$ are the atomic mass and the electron mass, respectively.
We calculated $\Delta E_\text{R}$'s to be 0.051 eV (XPS), 0.185 eV (HXPS), and 0.272 eV (HX-PES). 
Accordingly the calculated shifts in the binding energy of the O $1s$ peak with respect to the XPS
spectrum are 0.134 eV (HXPS) and 0.221 eV (HX-PES).
Considering the energy resolution, these results are in good agreement with the observed values.

\begin{figure}[t]
	\centering
	\includegraphics[width=86mm]{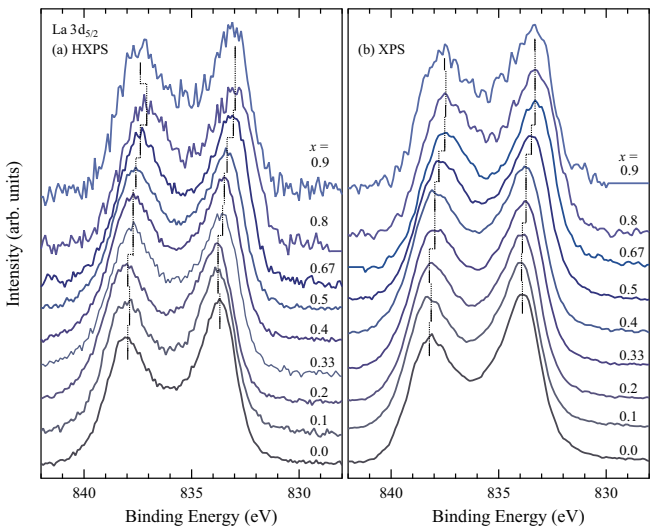}
	\caption{
		La $3d_{5/2}$ core-level photoelectron spectra of LSMO for various Sr concentrations,
		which were acquired by (a) HXPS and (b) XPS.
	}
	\label{fig3}
\end{figure}

Figures \ref{fig3}(a) and \ref{fig3}(b) show La $3d_{5/2}$ core-level spectra of LSMO in the
whole range of $x$, as measured by HXPS and XPS, respectively.
The HXPS peak at near 833 eV [Fig.\ \ref{fig3}(a)] is narrower and more symmetric than the
corresponding XPS peak [Fig.\ \ref{fig3}(b)].
Considering the comparable energy resolutions available from HXPS and XPS and the different
surface sensitivity resulting from the different kinetic energy of photoelectrons ($\sim$4570 eV for
HXPS against $\sim$660 eV for XPS), this result likely indicates that surface contributions at
the higher-binding-energy side of the peak are suppressed in the HXPS measurement.

\begin{figure}[t]
	\centering
	\includegraphics[width=86mm]{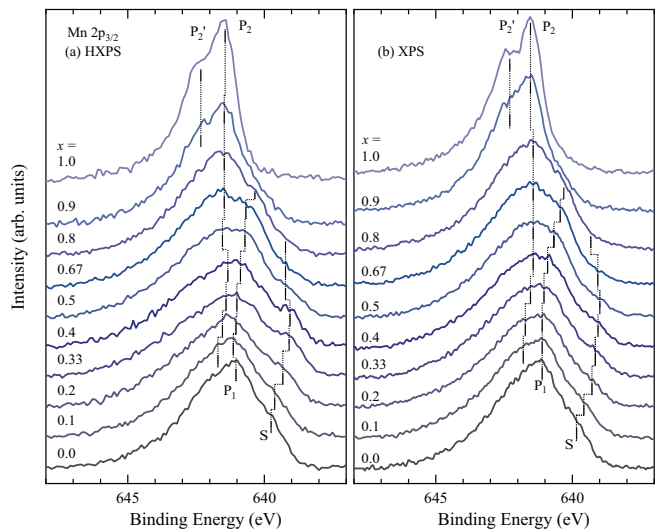}
	\caption{
		Mn $2p_{3/2}$ core-level (a) HXPS and (b) XPS spectra of LSMO for various Sr
		concentrations $x$.
		S denotes the well-screened final state at the Mn$^{3+}$ site.\cite{ref22}
		P$_1$ corresponds primarily to the main peak of Mn$^{3+}$.
		P$_2$ and P$^\prime_2$ are assigned to the main peak of Mn$^{4+}$.
	}
	\label{fig4}
\end{figure}

\begin{figure}[t]
	\centering
	\includegraphics[width=75mm]{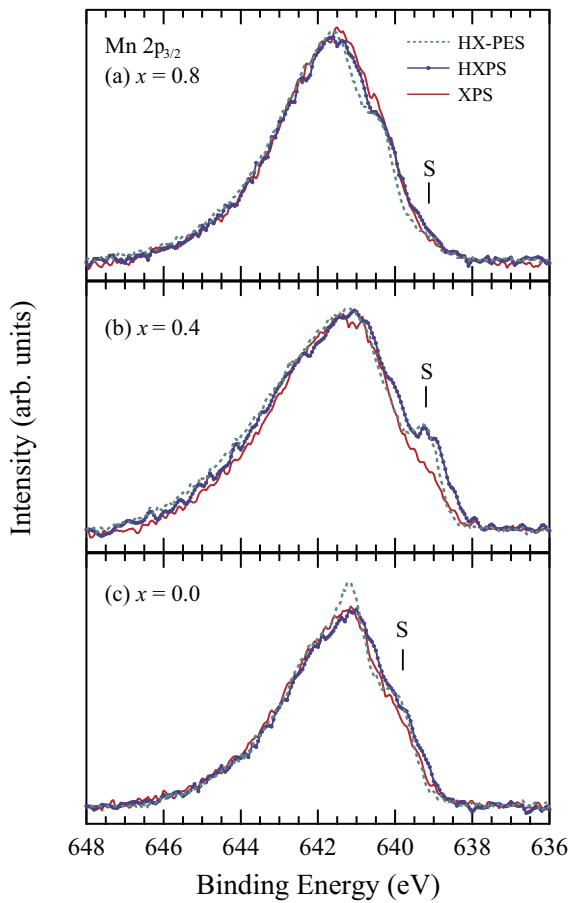}
	\caption{
		Comparison of Mn $2p_{3/2}$ core-level photoelectron spectra of LSMO measured
		by HX-PES, HXPS, and XPS.
		(a) $x=0.0$, (b) $x=0.4$, and (c) $x=0.8$.
	}
	\label{fig5}
\end{figure}

Figures \ref{fig4}(a) and \ref{fig4}(b) show Mn $2p_{3/2}$ core-level spectra of LSMO
acquired by HXPS and XPS, respectively.
These spectra show that two or three features change systematically with $x$.
The low-energy shoulder labeled S is due to the wellscreened final state of Mn$^{3+}$.\cite{ref22}
The lower-energy peak P$_1$ is primarily due to the main peak of Mn$^{3+}$, and the two
higher-energy features P$_2$ and P$^\prime_2$ are due to the main peak of
Mn$^{4+}$.\cite{ref23,ref24}
The peak positions and the line shapes are the same for the HXPS and XPS spectra, but the
shoulder feature S differs, as shown more clearly in Fig.\ \ref{fig5}.

Figure \ref{fig5} compares the Mn $2p_{3/2}$ core-level spectra measured by HX-PES, HXPS,
and XPS.
Only for $x = 0.4$ does the intensity of the shoulder S in the XPS spectrum differ significantly
from that obtained by HX-PES or HXPS.
This result comes from the combination of two facts: the LSMO surface is less conductive than
the bulk and XPS is less bulksensitive than HX-PES and HXPS.\cite{ref23,ref25}
For $x = 0.0$, the peak in the HX-PES spectrum is shaper than in the HXPS or XPS spectra
[Fig.\ \ref{fig5}(a)] and for $x = 0.8$, the lineshape of the HX-PES spectrum is narrower than
that of the HXPS or XPS spectra [Fig.\ \ref{fig5}(c)], both of which we attribute to the
better energy resolution of the HX-PES measurement.
Thus the comparisons in Figs.\ \ref{fig4} and \ref{fig5} show clearly that HXPS is
as bulk-sensitive as HX-PES and has a practical throughput comparable to that of XPS.

\begin{figure}[t]
	\centering
	\includegraphics[width=75mm]{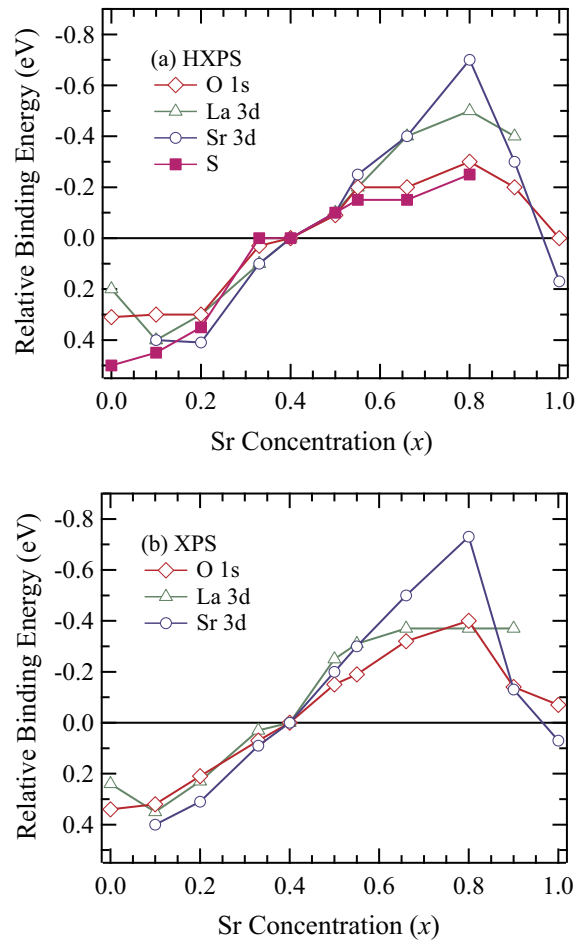}
	\caption{
		Core-level binding-energy shift ($\Delta E_\text{B}$) of LSMO plotted as a
		function of Sr concentration $x$ deduced from (a) HXPS and (b) XPS.
		$x=0.4$ is chosen as reference.
	}
	\label{fig6}
\end{figure}

Figure \ref{fig6} shows the core-level binding-energy shifts $\Delta E_\text{B}$ of the La
$3d_{5/2}$, Sr $3d_{5/2}$, and O $1s$ core levels together with the shift of the shoulder
structure of the Mn $2p_{3/2}$ (S in Fig.\ \ref{fig4}) as a function of $x$.
To determine $\Delta E_\text{B}$, we fit the corelevel spectra with multiple Voigt functions.
Figure \ref{fig6} shows that the behavior of $\Delta E_\text{B}$ for the La $3d_{5/2}$,
Sr $3d_{5/2}$, and O $1s$ core levels are basically identical for $0.0 \leq x \leq 0.67$ regardless
of whether the spectra were acquired by XPS or by HXPS.
This behavior is also in good agreement with that reported for $0.0 \leq x \leq 0.6$.\cite{ref26}
Figures \ref{fig1} and \ref{fig3} show that the increase in $\Delta E_\text{B}$ with $x$ continues
up to $x = 0.8$.
In contrast, an obvious discontinuity in $\Delta E_\text{B}$ is apparent from the spectra
above $x = 0.8$, where $\Delta E_\text{B}$ decreases rapidly.

$\Delta E_\text{B}$ can theoretically be described as \cite{ref16,ref26,ref27,ref28,ref29}
\begin{equation*}
	\Delta E_\text{B} = \Delta\mu - K\Delta Q + \Delta V_\text{M} - \Delta E_\text{R},
\end{equation*}
where $\Delta\mu$ is the chemical-potential shift as a function of $x$, $\Delta Q$ is the change
in the number of valence electrons of the relevant ion, $K$ is an empirical constant,
$\Delta V_\text{M}$ is the change in the Madelung potential at the relevant ion, and
$\Delta E_\text{R}$ is the change in the inter-atomic relaxation energy of the core hole due to
screening by metallic conduction electrons.
The $K\Delta Q$ term is zero for these three core levels.
No obvious discontinuity occurs across the metal–insulator transition between $x = 0.1$ and 0.2,
which implies that $\Delta E_\text{R}$ is negligible.
Therefore, we expect
\begin{equation*}
	\Delta E_\text{B} \approx \Delta \mu + \Delta V_\text{M}.
\end{equation*}
Theoretical and experimental studies of $\Delta E_\text{B}$ established that $\Delta V_\text{M}$
is negligible for typical carrier-doped systems such as La$_{1-x}$Sr$_x$MnO$_3$ or
La$_{2-x}$Sr$_x$CuO$_4$ \textit{within the same crystal structure}.
Thus, we conclude that $\Delta E_\text{B} \approx \Delta \mu$ for $0.0 \leq x \leq 0.8$.
\cite{ref16,ref26,ref27,ref28,ref29}
On the other hand, the chemical potential of these hole-doped systems should decrease continuously
with $x$ (namely, $\Delta\mu$ decreases with $x$) unless a large gap suddenly opens.
Therefore, the observed rapid turn or discontinuity of $\Delta E_\text{B}$ at $x = 0.8$ is most likely
due to a change in $\Delta V_\text{M}$ caused by a structural phase transition from a (pseudo)
cubic phase ($0.0 \leq x \leq 0.8$) to a hexagonal phase ($0.9 \leq x$).\cite{ref23}

Noted that $\Delta E_\text{B}$ for the feature S in the Mn $2p_{3/2}$ peak shows exactly
the same behavior as $\Delta E_\text{B}$ for the other three core levels, particularly the O $1s$
HXPS core level whereas the other components P$_1$, P$_2$, and P$^\prime_2$ have characteristics
different from those of S (see Fig.\ \ref{fig4}).
So far it has been reported that $\Delta E_\text{B}$ for Mn $2p_{3/2}$ is distinctly different from
$\Delta E_\text{B}$ for the above three core levels.\cite{ref16,ref26}
The result of these reports is attributed to their analysis that did not (actually could not) decompose
the Mn $2p_{3/2}$ peak into multiple components although the average valence is changing from
$3+$ to $4+$ (i.e., $\Delta Q \neq 0$).
In the present analysis, the high bulk sensitivity and large practical throughput of HXPS enables us
to extract for the first time a component of $\Delta E_\text{B}$ proportional to $\Delta\mu$ from the
Mn $2p_{3/2}$ peak.
The behavior of feature S may be explained rather simply: S is the well-screened Mn $2p_{3/2}$
final state at the Mn$^{3+}$ site and the chemical potential $E_\text{F}$ is in the $e_g$ band
(except for $x \approx 1.0$), which means that $E_\text{F}$ is ``located at the Mn$^{3+}$''
that have $e_g$ electrons.

\section{Conclusions}

By using XPS, HXPS, and HX-PES, we have performed a comparative study of the electronic structure
of the CMR manganite La$_{1−x}$Sr$_x$MnO$_3$ for the whole range of $x$.
We observed the recoil effect in the O $1s$ HXPS spectra, which confirms theoretical predictions.
No appreciable surface contribution appears in the O $1s$ HXPS spectra, and an enhanced
low-energy shoulder structure appears in the Mn $2p_{3/2}$ HXPS spectra.
By combining high bulk sensitivity and large practical throughput, HXPS enables us to track
the lowenergy shoulder structure in the Mn $2p_{3/2}$ spectra over a wide range of $x$,
which gives new insight into the binding-energy shift of the Mn $2p_{3/2}$ core level.
The throughput and the energy resolution of HXPS is thus as good as that of XPS, and the bulk
sensitivity of HXPS is as good as that of HXPES.
Together, the results demonstrate that laboratory-based HXPS has significant potential for
general or systematic analysis of bulk electronic structure.

\begin{acknowledgments}
We are grateful to H.\ Kozuka for providing us with the LSMO samples used in this work.
The synchrotron radiation experiments at SPring-8 were performed under the approval
of the Japan Synchrotron Radiation Research Institute
(Proposals No.\ 2011A1420 and No.\ 2011B1710).
\end{acknowledgments}


\begin{thebibliography}{99}
\bibitem{ref24} The spectral weight of the Mn$^{4+}$ well-screened final state is hidden behind
	the Mn$^{3+}$ main peak. See Ref.\ \onlinecite{ref23}.
\end{thebibliography}
\end{document}